\documentclass[amsmath,twocolumn]{aastex702}

\usepackage{upgreek}
\usepackage{amssymb}
\usepackage{placeins}

\begin{document}

\title{The reddening of NGC 7469 and evidence for variable extinction}

\author[orcid=0000-0003-4888-2009,sname='Gaskell']{C. Martin Gaskell}
\affiliation{Department of Astronomy \& Astrophysics, University of California Santa Cruz}
\email[show]{mgaskell@ucsc.edu}  

\author[sname='Kamath']{Ananya R. Kamath} 
\affiliation{Department of Astronomy \& Astrophysics, University of California Santa Cruz}
\affiliation{Department of Astronomy, University of California Berkeley}
\email{Ananyark@berkeley.edu}

\author[gname=Emily,sname=Kwan]{Emily I. Kwan}
\affiliation{Department of Astronomy \& Astrophysics, University of California Santa Cruz}
\affiliation{College of Computing, Georgia Institute of Technology}
\email{emilykwan999@gmail.com}

\author[gname=Divya,sname=Subramonian]{Divya G. Subramonian}
\affiliation{Department of Astronomy \& Astrophysics, University of California Santa Cruz}
\affiliation{Department of Computer Science, University of California Santa Barbara}
\email{divya.subramonian25@gmail.com}


\begin{abstract}

We estimate the reddening of the broad-line region and continuum of the active galactic nucleus NGC~7469 during the 1996 {\it International AGN Watch} multi-wavelength monitoring campaign using seven different reddening indicators.  As was found for NGC~5548, these indicators support velocity-integrated broad hydrogen line ratios being close to Case B values. All the indicators point to substantial reddening of $E(B-V) = $ 0.44 $\pm$ 0.03 for NGC~7469 during mid-1996. We find evidence for a gradual increase of about 10\% in the reddening during the seven weeks of the 1996 campaign.  Decades-long optical monitoring before and after the campaign is also consistent with modest changes in the extinction from year to year.

\end{abstract}




\section{Introduction} 

Knowing the internal reddenings of active galactic nuclei (AGNs) is crucial for understanding AGNs.  If reddening is neglected one obtains incorrect line ratios and continuum shapes, and underestimates the luminosity.  Although it has been recognized for decades that there is substantial reddening of AGNs (see review by \citealt{Gaskell17}), it has remained common practice to only correct for local Galactic reddening and to assume that internal reddening in the AGNs and their host galaxies is negligible. Neglecting internal reddening leaves one with optical continuum shapes that differ from what is expected theoretically and which vary from object-to-object.  Accretion disk sizes calculated from the uncorrected luminosities are significantly smaller than sizes implied by reverberation mapping and microlensing studies, a conflict which can be resolved if reddening is allowed for \citep{Gaskell17}.

\citet{Gaskell+23}(G23) have estimated the reddening of the best-observed AGN, NGC~5548, using seven different and mostly independent reddening indicators covering the IR to the UV.  They have shown that the different indicators consistently give a total reddening of $E(B-V) = 0.25 \pm 0.03$ for NGC 5548 when it was being simultaneously monitored in the UV and optical by the {\it International AGN Watch} for 10 weeks in 1993 (see \citealt{Korista+95}). G23 assumed that the three hydrogen line ratios (Ly$\upalpha$/H$\upbeta$, H$\upalpha$/H$\upbeta$, and Pa$\upbeta$/H$\upbeta$) had Case B ratios. They found that the reddenings deduced from the hydrogen line ratios agreed with the non-hydrogenic reddening indicators, thus supporting the assumption of Case B hydrogen line ratios for the velocity-integrated flux ratios.

In this paper we consider a second well-studied AGN, NGC~7469, which was intensively monitored in the X-ray, UV and optical bands by the {\it International AGN Watch} in mid-1996.  Our goal is to see whether it too has substantial reddening and to see if the different reddening indicators again give consistent results.  As far as possible, we conduct a similar analysis to the G23 analysis of NGC~5548.  

For the 10-week period in 1993 when NGC~5548 was being simultaneously monitored in the UV and optical G23 found that the reddening was constant, but outside this period the H$\upalpha$/H$\upbeta$ ratio has shown substantial variation \citep{Shapovalova+04}.  If the H$\upalpha$/H$\upbeta$ ratio is an indicator of reddening, variability of the ratio therefore implies variability of the reddening.  Balmer-line profiles of AGNs show changes in asymmetry on timescales longer than the light-crossing time.  These profile variations are not correlated with continuum variability and are on a longer timescale than continuum variability \citep{Wanders+Peterson96}.  \citet{Gaskell+Harrington18} have demonstrated that patchy dust moving across the line of sight can explain variation of the Balmer line profiles of NGC~5548. Given the possibility of variable extinction, we look at reddening indicators for NGC~7469 as a function of time where possible to investigate the possibility of reddening variation.

To obtain $E(B-V)$ from reddenings determined from wavelengths other than those of the $B$ and $V$ passbands it is necessary to assume a reddening curve.  \citet{Gaskell+23} considered three different reddening curves: a standard Milky Way curve, a steeply-rising, SMC-like curve, and the mean AGN reddening curve of \citet{Gaskell+Benker07}(GB07).  They found that the average reddening of NGC~5548 was similar for all three reddening curves, but the GB07 curve gave the most consistent results across different wavelength regions.  In particular, a steep SMC-like reddening curve gave $E(B-V)$ values that were too low in the UV and too high in the IR.  In this paper we therefore only use the GB07 reddening curve.  The conversion factors to go from reddenings between the various passbands and $E(B-V)$ are given in Table 1 of G23. We adopt the same intrinsic ratios of the reddening indicators as G23.  These are discussed in G23 and summarized in Table 1 there.

\section{OBSERVATIONS}
\label{sec:observations}

The {\it International AGN Watch} monitored NGC 7469 in mid-1996. The AGN was observed by the {\it International Ultraviolet Explorer} (IUE) from 1996 June 10 to July 29 \citep{Wanders+97}. In parallel with this, 200 ground-based spectra were acquired and carefully put on a homogeneous scale as described in \citep{Collier+98}. During this monitoring campaign, a single high signal-to-noise UV spectrum covering 1150–3300 Å was also obtained using the {\it Faint Object Spectrograph} (FOS) on the {\it Hubble Space Telescope} (HST) on 1996 June 18 \citep{Kriss+00}.


\section{He II \texorpdfstring{$\uplambda$}{}1640/\texorpdfstring{$\uplambda$}{}4686}

The \ion{He}{2} $\uplambda$1640 and $\uplambda$4686 lines have long been recognized as a potentially valuable reddening indicator for AGNs \citep{Shuder+MacAlpine79,MacAlpine81}. However, in practice the ratio has not been used much because both lines are strongly blended with other lines.  \ion{He}{2} $\uplambda$1640 is blended with \ion{O}{3}] $\uplambda$1663 and \ion{He}{2} $\uplambda$4686 is blended with an infestation of broad \ion{Fe}{2} emission that frequently makes the \ion{He}{2} $\uplambda$4686 line unrecognizable.  G23 have demonstrated that the blending problems can be overcome by finding the $\uplambda$1640/$\uplambda$4686 ratio from difference spectra since \ion{He}{2} is much more variable than the contaminating lines.  We follow a similar procedure here. 

\citet{Wanders+97} and \citet{Kriss+00} present measurements of the UV \ion{He}{2} $\uplambda$1640 line from the large number of {\it IUE} spectra during the 1996 campaign.  Unfortunately, the {\it IUE} spectra are far inferior to the {\it HST} spectra used by G23 to create a \ion{He}{2} $\uplambda$1640 difference spectrum for NGC~5548.  From the \ion{He}{2} $\uplambda$1640 light curves we were able to identify a \ion{He}{2} high state and a low state which had Crimean Astrophysical Observatory (CrAO) spectra within $\pm 2$ days of {\it IUE} spectra. The high state had CrAO spectra taken on JDs 2450245.5 and 2450246.5; the low state had spectra taken on JDs 2450283.5, 2450286.5 and 2450286.5  For the high and low states we averaged these and averaged the {\it IUE} spectra taken within $\pm 2$ days.  The continuum-subtracted \ion{He}{2} profiles are shown in Fig.~1.  The two spectra are plotted on the same scale and the area under the curves is proportional to the lines' energy.

\begin{figure} 
	\begin{center}
		\includegraphics[width = \columnwidth]{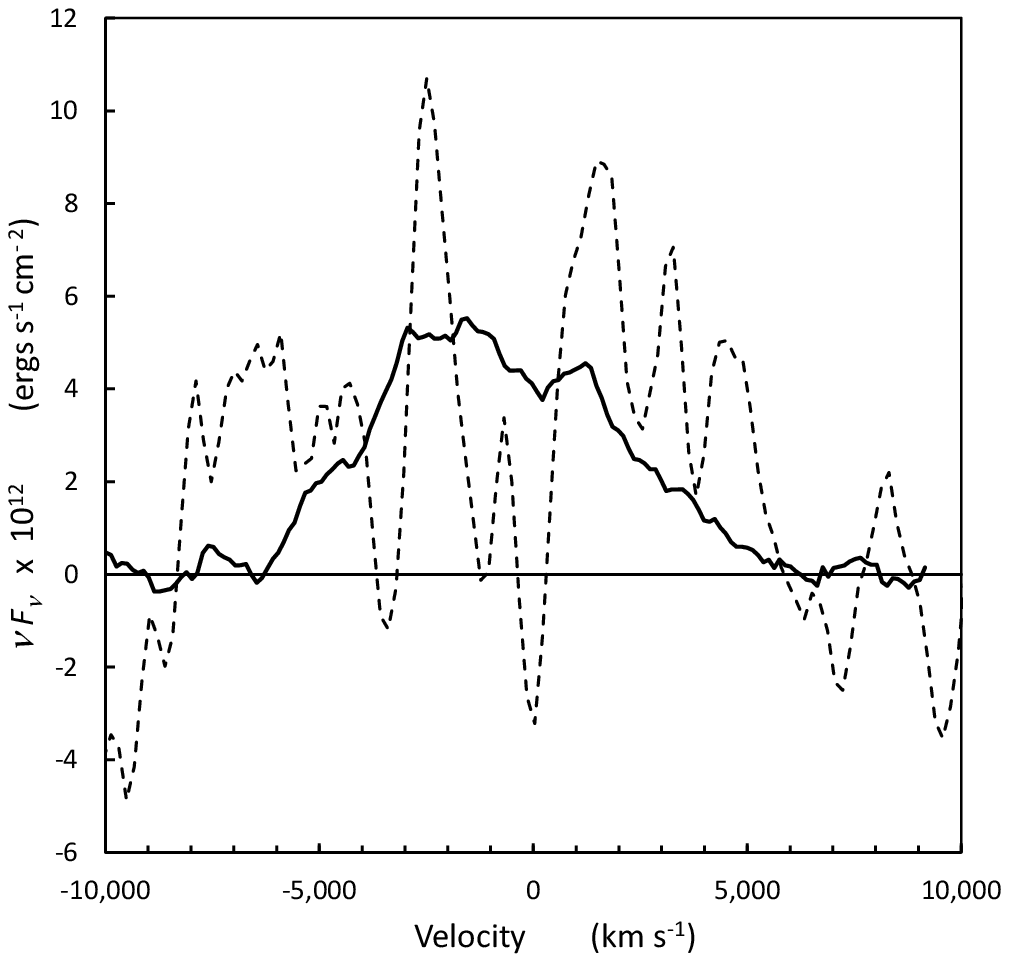}
        \caption{Continuum-subtracted difference spectra for \ion{He}{2} $\uplambda$1640 and $\uplambda$4686. $\uplambda$4686 is shown as a solid line and $\uplambda$1640 as a dashed line. The spectra are JD 2450245.5 minus the average of the low state from JD 2450283.5 to JD 2450286.5. See text for details.}  
	\end{center}
\end{figure}

It can be seen that, while the profile of \ion{He}{2} $\uplambda$4686 is well defined and typical of broad-line profiles, the profile of \ion{He}{2} $\uplambda$1640 is very noisy.  The $\uplambda$1640/$\uplambda$4686 ratio is 1.4 which gives $E(\uplambda1640 - \uplambda4686) = 2.0$.  The main source of error is in the setting of the continuum level.  The lowest the continuum could be set is about $2 \times 10^{-12}$ ergs s$^{-1}$ cm$^{-2}$ lower, which gives a conservative upper limit to $\uplambda$1640/$\uplambda$4686 ratio of 2 corresponding to $E(\uplambda1640 - \uplambda4686) = 1.6$.  We therefore consider $E(\uplambda1640 - \uplambda4686)$ to be uncertain by $\pm 0.4$.  Adopting $E(\uplambda 1640 - \uplambda 4686)/E(B-V) = 4.26$ from \citet{Gaskell+Benker07} gives $E(B-V) = 0.46 \pm 0.09$.




\section{O I \texorpdfstring{$\uplambda$}{}1304/\texorpdfstring{$\uplambda$}{}8446}
\label{sec:OI}

The \ion{O}{1} $\uplambda$8446 and $\uplambda$1304 lines mostly arise in a direct cascade (see Fig.~1 of \citealt{Grandi83}), so the theoretical ratio is expected to be the ratio of the wavelengths: i.e., $8446/1306 = 6.48$.  The {\it Hubble Space Telescope} spectrum of NGC~7469 taken on JD 2,450,253 during the {\it International AGN Watch} campaign gives an \ion{O}{1} $\uplambda$1304 flux of $2.1 \times 10^{-13} \pm 0.16 \times 10^{-13}$ ergs s$^{-1}$ cm$^{-2}$ \citep{Kriss+00}. However, the average flux from the {\it IUE} monitoring for the whole campaign was $1.4 \times 10^{-13} \times 10^{-13}$ ergs s$^{-1}$ cm$^{-2}$.

To our knowledge, \ion{O}{1} $\uplambda$8446 was not observed during the campaign, but it was observed a decade later by \citet{Landt+08} in January of 2006.  Because the observations were far from simultaneous, we scale the \ion{O}{1} $\uplambda$8446 flux to the H$\beta$ line.  Our measurements of the optical and IR spectra taken by \citet{Landt+08} give \ion{O}{1} $\uplambda$8446/H$\upbeta = 0.24 \pm 0.02$ which is typical of other AGNs they observe.\footnote{Note that in the case of NGC~7469, the published IR line fluxes in \citet{Landt+08} are too low by a factor of 2.6.  We are grateful to Hermine Landt for providing us with the original spectra to resolve this discrepancy.}

Around 1996 June 18 (= JD 2450252.5) when the {\it HST} spectrum of NGC~7469 was taken, the mean H$\upbeta$ flux was $8.25 \times 10^{-13}$. If we assume the \ion{O}{1} $\uplambda$1304/$\uplambda$8446 ratio stays constant, then the observed H$\upbeta$ flux implies that the \ion{O}{1} $\uplambda$8446 flux was $2.0 \times 10^{-13}$. This gives an observed $\uplambda$1304/$\uplambda$8446 ratio of 1.05. Assuming that the intrinsic \ion{O}{1} $\uplambda$1304/$\uplambda$8446 ratio is 6.5, this gives $E(\uplambda1304 - \uplambda8446)$ = 1.69.  For the GB07 reddening curve this gives $E(B-V) = E(\uplambda1304 - \uplambda8446) /6.75 = 0.29$.  If we use the {\it IUE} flux for the whole campaign we get $E(B-V)  = 0.36$ instead.


\section{The Optical Continuum}

\citet{Choloniewski81} made the important discovery that there is a linear relationship between optical fluxes at different wavelengths.  This implies that the spectral shape of the variable continuum stays the same but there is an additional constant contribution at each wavelength due to non-varying starlight and extend line emission. Cho\l{}oniewski postulated that different observed slopes in flux-flux plots for different AGNs are due to different reddenings. For an externally-illuminated accretion disk, \citet{Friedjung85} showed that the spectral shape is $F_{\nu} \propto \nu^{+1/3}$ at optical wavelengths. Observationally, \citet{Heard+Gaskell23} find the optical--UV continuum shape of the bluest AGNs in a large SDSS sample to be consistent with this. 

To calculate $E(B-V)$ using the Cho\l{}oniewski method, the slope of the observed fluxes between given wavelengths is compared with the expected ratio. If the flux from an accretion disk varies as $F$ $\upalpha$ $\nu^{+1/3}$, the expected slope in the flux-flux plot would be the ratio of the two given wavelengths to the power of +1/3. In Fig.~2, we plot fluxes from $\uplambda$4845 and $\uplambda$6962 from \citet{Collier+98}.  Because the errors in the two fluxes are comparable, we follow \citet{Heard+Gaskell23} and use the least squares bisector method of \citet{Isobe+90} to estimate the slope.  We get a slope of $0.61 \pm 0.14$.  

For $F$ $\upalpha$ $\nu^{+1/3}$ we expect the intrinsic ratio $F_{4845}/F_{6962}$ to be $(6962/4845)^{+1/3} = 1.128$. The observed slope of 0.61 gives a flux ratio of $(6962/4845)^{-0.61} = 0.802$. This corresponds to $E(\uplambda4845 - \uplambda6962) = 0.67 \pm 0.17$. Adopting $E(\uplambda4845 - \uplambda6962)/E(B-V) = 1.44$ from GB07 gives $E(B-V) = 0.46 \pm 0.12$.

\begin{figure} 
	\begin{center}
		\includegraphics[width = 1.0\linewidth]{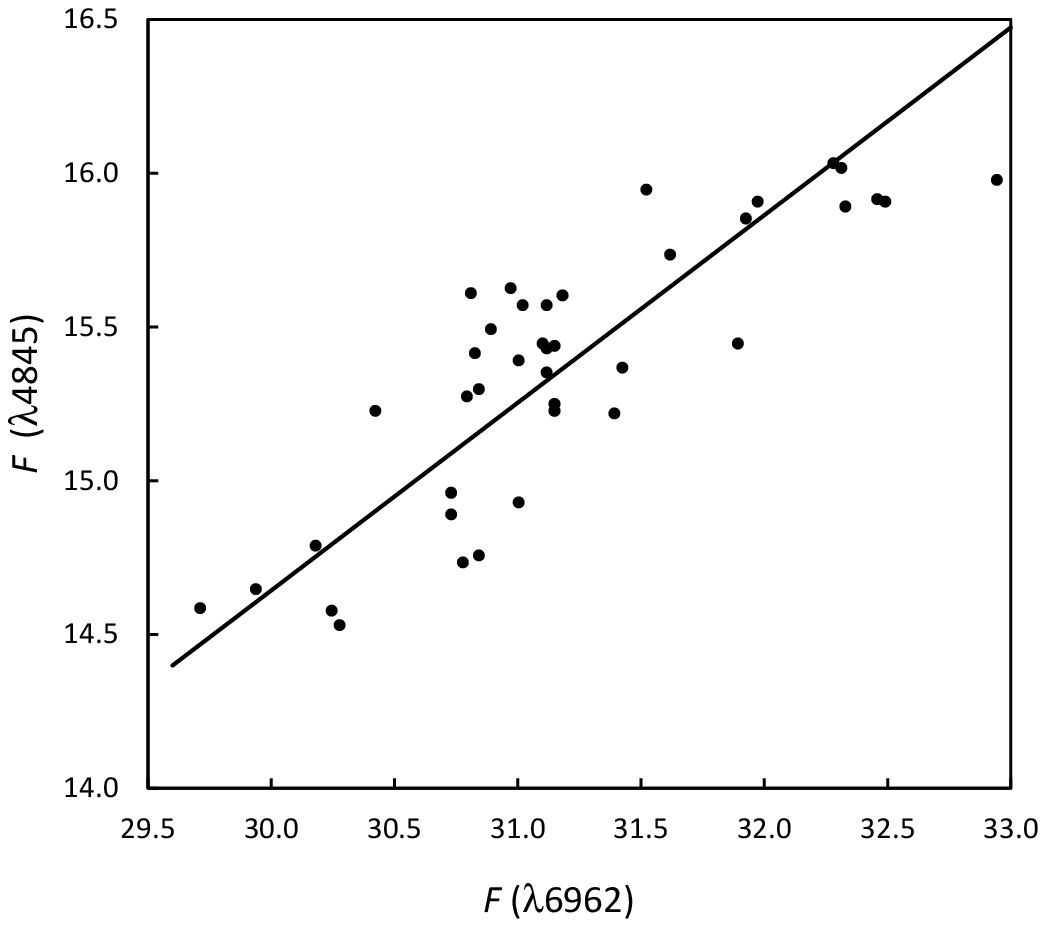}
        \caption{Continuum fluxes for NGC~7469 in mJy at $\uplambda$4845 versus fluxes at $\uplambda$6962. The line is a least-squares bisector fit \citep{Isobe+90}}. 
	\end{center}
\end{figure}

\section{UV-to-optical continuum}

\begin{figure} 
	\begin{center}
		\includegraphics[width = 1.0\linewidth]{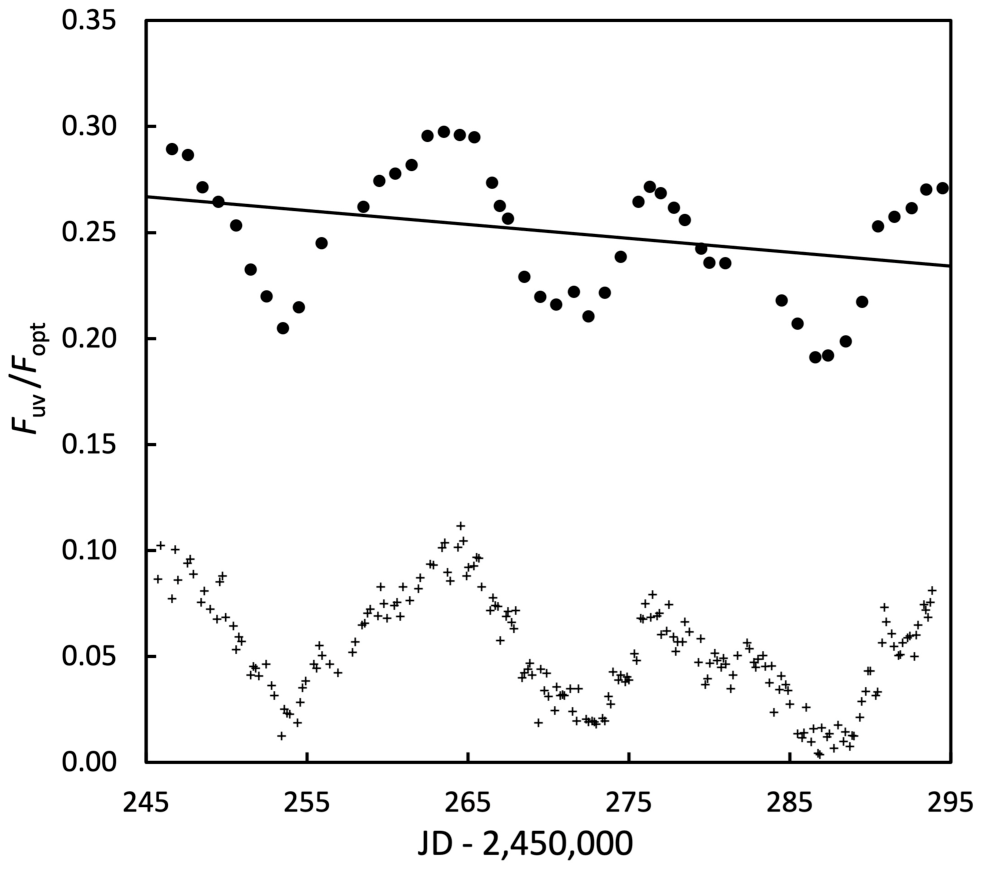}
        \caption{The filled circles show the ratio of near simultaneous $\uplambda$1315 and starlight-subtracted $\uplambda$4865 fluxes as a function of time for NGC~7469. The line is a least squares fit to indicate a possible long-term trend. For comparison, the $\uplambda$1315 fluxes with an arbitrary offset and scaling are shown as small crosses at the bottom of the plot. }  
	\end{center}
\end{figure}

In Fig.~3 we plot the UV-to-optical continuum ratio after removing the contribution of starlight to the optical flux. We use UV fluxes at $\uplambda$1315 given by \citet{Kriss+00} and optical fluxes at $\uplambda$4865 from \citet{Collier+98} with the starlight contribution given in Table 12 of \citet{Bentz+13} subtracted. Both sets of these observations were taken $\pm 2$ days within each other. For comparison, we also show arbitrarily-scaled UV $\uplambda$1315 flux. It can be seen that the UV-to-optical ratio changes on a timescale of days. This is because, on this timescale, the UV varies more than the optical. The average ratio over the course of the monitoring is 0.25. The maximum is 0.30, and the minimum is 0.19. Therefore, we adopt an uncertainty of $\pm 0.05$.  If we assume that the $F_{\nu} \propto \nu^{+1/3}$ spectrum of an externally-illuminated accretion disk applies for the range of wavelengths we consider, then the theoretical $F_{\lambda1315}/F_{\lambda4865}$ ratio is 1.36. Using the GB07 reddening curve of gives $E(B-V)$ to be 0.39 $\pm0.04$. 


\section{Hydrogen Line Ratios }
\label{sec:hydrogen}

\subsection{H\texorpdfstring{$\upalpha$}{alpha}/H\texorpdfstring{$\upbeta$}{}}
\label{sec:Ha/Hb}

\citet{Collier+98} give H$\upalpha$ and H$\upbeta$ fluxes for the 1996 monitoring period.  They find that the Balmer lines lag the $\uplambda$1315 continuum by $\sim 5 - 6$ days.  In Fig.~4 we show the H$\upalpha$/H$\upbeta$ ratio calculated from near-simultaneous observations as a function of time. It can be seen that there is a gradual, highly-significant increase in the ratio over the two-month monitoring period. The H$\upalpha$/H$\upbeta$ ratio increases from 4.7 to 5.2, which, for a Case B unreddened ratio implies reddenings increasing from $E(B-V) = 0.46$ to $0.55$.  It can be seen in Fig.~4 that there is no correlation with the UV or X-ray flux.  We discuss the implications of the gradual increase in Section 6.

\subsection{Lyman \texorpdfstring{$\upalpha$}{}/Balmer Lines}
\label{Lya/Balmer}

\citet{Wanders+97} and \citet{Kriss+00} give Ly$\upalpha$ fluxes for the monitoring period.  Fig.~5 shows the ratios of near-simultaneous observations of Ly$\upalpha$, H$\upbeta$, and H$\upalpha$. It can be seen in the top panel that the Ly$\upalpha$/H$\upalpha$ ratio shows a gradual decrease. A decrease would be expected if the increase in the H$\upalpha$/H$\beta$ ratio in Fig~4 were just due to H$\upalpha$ increasing.  However, the Ly$\upalpha$/H$\upbeta$ ratio (lower panel) also shows a decline with time that would not be expected if the increase in Ly$\upalpha$/H$\upbeta$ were due to just H$\upbeta$ decreasing.  The simplest explanation is that the reddening was gradually increasing during the monitoring campaign.

If, following G23, we assume intrinsic Ly$\upalpha$/H$\upbeta$ and Ly$\upalpha$/H$\upalpha$ ratios of 35 and 12 respectively (see \citealt{Gaskell17}), the decrease in Ly$\upalpha$/H$\upbeta$ from approximately 7.5 to 4.8 over the period of monitoring implies an increase in reddening from $E(B-V) = 0.34$ to $0.44$, and the decrease in Ly$\upalpha$/H$\upalpha$ from  1.33 to 1.0 implies an increase in reddening from $E(B-V) = 0.49$ to $0.55$.  We discuss the significance of the changes in Section 6.

\begin{figure} 
	\begin{center}
		\includegraphics[width = 1.0\linewidth]{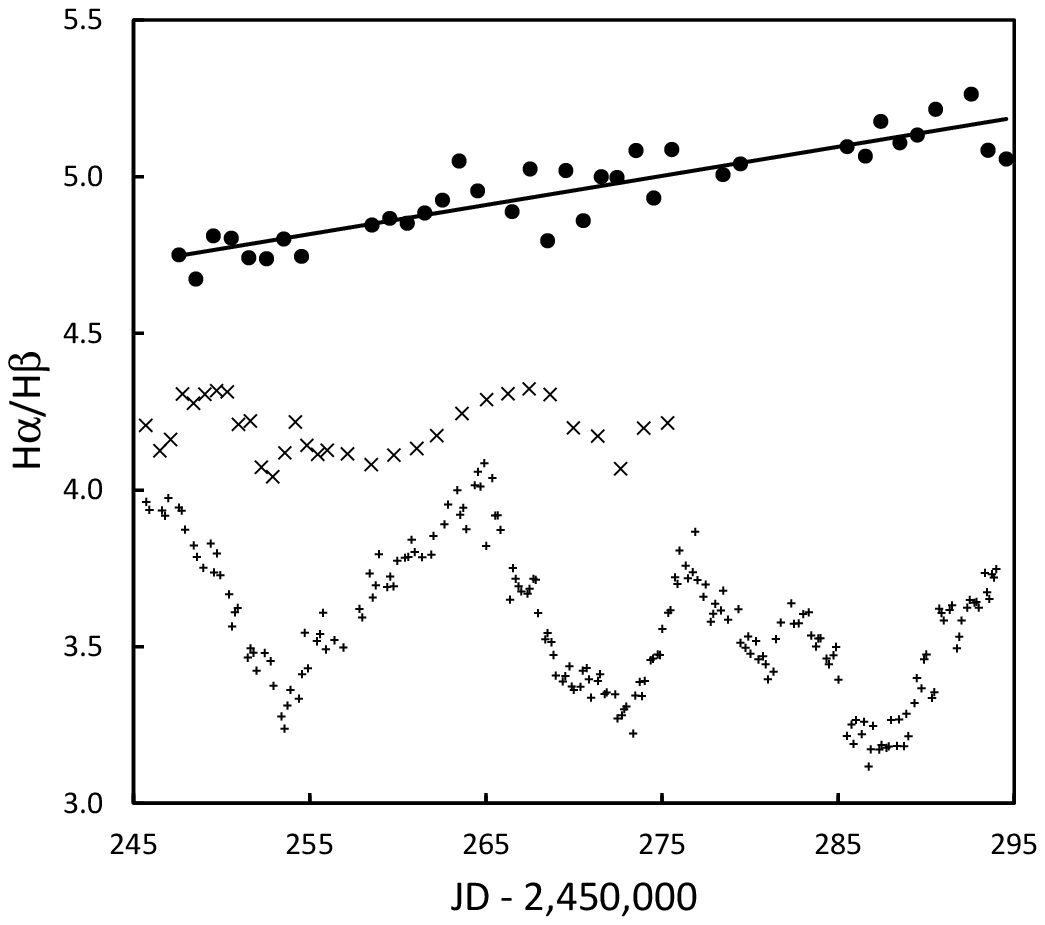}
        \caption{The filled circles at the top of the figure show the variation of the H$\upalpha$/H$\upbeta$ ratio for NGC~7469 as a function of time. The line is a least-squares fit. As in Fig.~3, the small vertical crosses (+) show the $\uplambda$1315 UV flux \citep{Wanders+97,Kriss+00} on an arbitrary scale.  The X's show average 2-10 keV fluxes as measured by {\it RXTE} \citep{Nandra+98}, also on an arbitrary scale.} 
	\end{center}
\end{figure}

\begin{figure}
\centering
    \begin{minipage}[b]{0.45\textwidth}
        \centering
        \includegraphics[width=\textwidth]{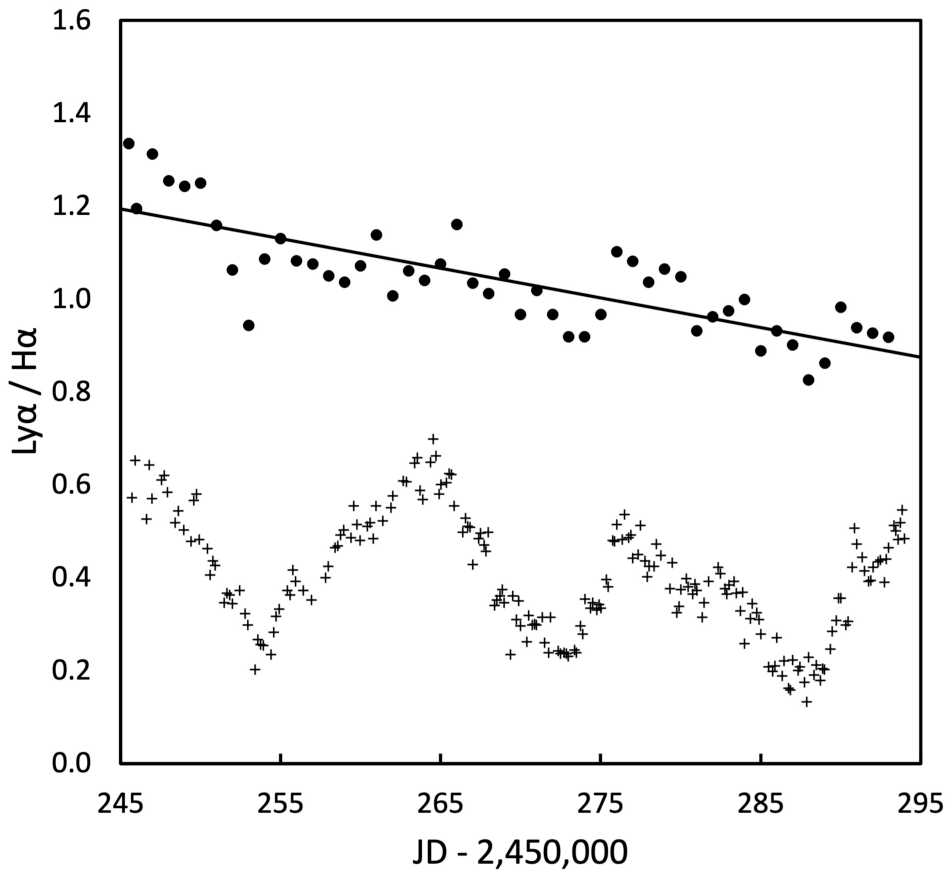}    
    \end{minipage}
    \hfill
    
    \begin{minipage}[b]{0.45\textwidth}
        \centering
        \includegraphics[width=0.9\textwidth]{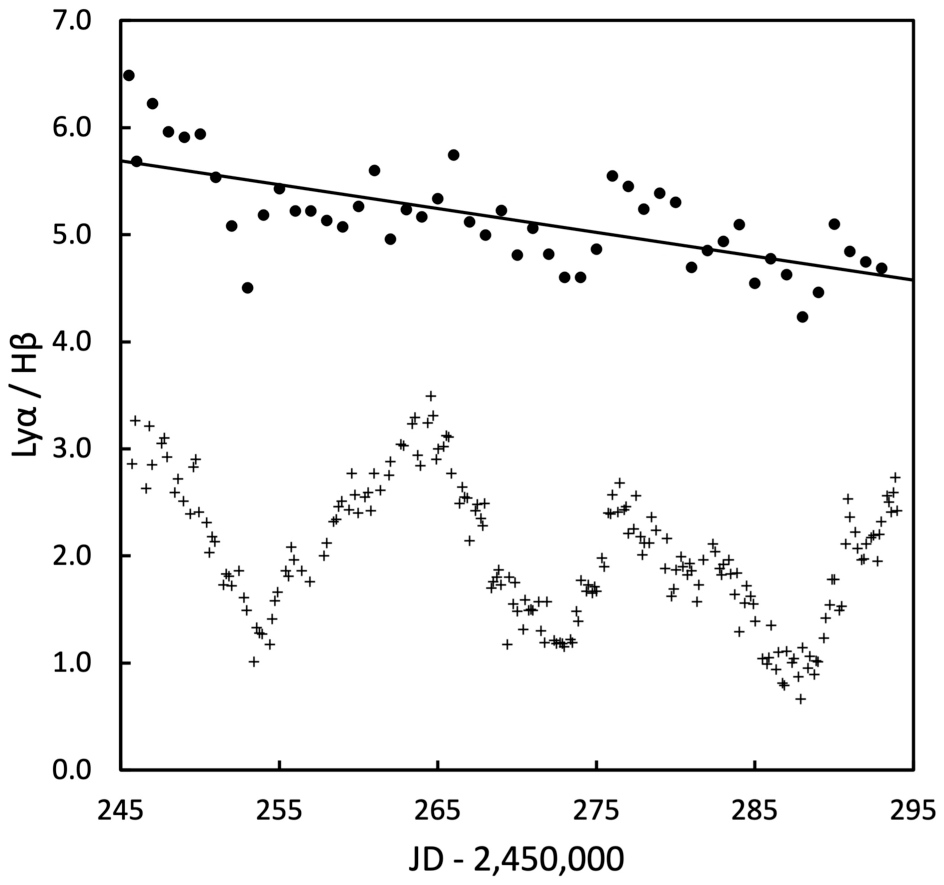}
    \end{minipage}
    \caption{The ratio of Ly$\upalpha$ to near-simultaneous Balmer lines fluxes for NGC~7469 as a function of time (filled solid circles).  The upper panel shows Ly$\upalpha$/H$\upalpha$ and the lower panel  Ly$\upalpha$/H$\upbeta$.  Balmer line fluxes are averages within $\pm3$ days of the UV observations. The lines are least squares fits. In both panels, the small crosses again show $\uplambda$1315 fluxes from \citep{Kriss+00} on an arbitrary scale for comparison.}
\end{figure}


\section{The Mean Reddening}
\label{sec:mean_reddening}

In Table 1, we summarize the $E(B-V)$ estimates given by each of the seven reddening indicators using the GB07 mean AGN reddening curve. Results are ordered by the shortest wavelength of each of the indicators. 

At the bottom of the table we give an unweighted mean of all the reddening estimates. Where there is a range of reddenings for an indicator, we useed the average. We get a mean reddening of $E(B-V)$ = 0.44 $\pm$ 0.03. The scatter of the individual estimates about this mean is consistent with the uncertainties of the estimates.   Each of the estimates gives a far higher reddening than the $E(B-V) = 0.017$ \citep{Schlafly+Finkbeiner11} caused by Galactic dust in solar neighborhood.  This indicates that the reddening we find arises almost entirely within the AGN itself.


\begin{table}
\begin{minipage}{81mm}
\caption{Reddening estimates for NGC~7469}
\label{symbols}
\centering
\begin{tabular}{@{}lcccc}
\hline																	
Feature(s)	&	$\uplambda_{\mathrm{short}}$	&	$\uplambda_{\mathrm{long}}$	&	$E(B-V)$				\\

\hline																	
Ly$\upalpha$/H$\upbeta$	&	$\uplambda$1216	&	$\uplambda$4861	&	$0.34 - 0.44$  \,\,  \\
Ly$\upalpha$/H$\upalpha$ &  $\uplambda$1216 &   $\uplambda$6563 &  $0.49 - 0.55$  \,\, \\
\ion{O}{1}	&	$\uplambda$1304	&	$\uplambda$8446	&	$0.29 - 0.36$	\,\,	\\
$F_{\uplambda1315}$/$F_{\uplambda4845}$	&	$\uplambda$1315	&	$\uplambda$4845	&	0.39	$\pm$	0.04	\\
\ion{He}{2}	&	$\uplambda$1640	&	$\uplambda$4686	&	0.46	$\pm$	0.09	\\
$F_{\uplambda4845}$/$F_{\uplambda6962}$	&	$\uplambda$4845	&	$\uplambda$6962	&	0.46	$\pm$	0.12	\\
H$\upalpha$/H$\upbeta$	&	$\uplambda$4861	&	$\uplambda$6563	&	~$0.46 - 0.55$  \,\, \\
	&		&		&						\\
Mean	&		&		&	0.44	$\pm$	0.03   \\
\hline																				
\end{tabular}
\end{minipage}
\end{table}


\section{Variability of the Reddening}
\label{sec:variability}

\subsection{Variability during the 1996 monitoring}

For NGC~7469 the H$\upalpha$/H$\upbeta$ ratio implies a gradual increase in $E(B-V)$ of 0.09 (see section 4.1), and the change in  the ratio of Ly$\upalpha$ to the Balmer lines (see section 4.2) implies an increase of 0.10 and 0.06 in $E(B-V)$ over the same time period. Taken together, these slow changes are consistent with a modest increase in reddening over the monitoring period in 1996.  An important question, however, is whether these changes in the observed hydrogen line ratios mean that the reddening was gradually changing, or whether the intrinsic (i.e., unreddened) hydrogen line ratios were changing.

Theoretically, the factors that can change intrinsic hydrogen line ratios are temperature, density, photon flux, turbulence, and optical-depth effects. At low densities, the Case B H$\upalpha$/H$\upalpha$ ratio decreases slightly with temperature \citep{Osterbrock+Ferland06}. Then, as the density
increases, the ratio decreases below the familiar low-density
Case B value of $\sim 2.8$ found under nebular conditions.
Higher densities (see Fig.~4 of \citealt{Gaskell17}) produce
flatter Balmer decrements. On the other hand, intrinsic H$\upalpha$/H$\upalpha$ ratios greater than Case B (i.e., steeper Balmer decrements) can only be produced through high optical depths. Although optical depth effects in hydrogen lines have been extensively studied theoretically over the past half century, they are unlikely to be relevant to real AGNs because the geometry of the BLR (see \citealt{Gaskell09} for a review) is such that photons have an easy escape out of the sides of clouds and leaving the BLR perpendicular to the accretion disk (see Fig.~3 of \citealt{Gaskell17}).

In AGNs, high photon fluxes, high temperatures, and high densities are expected as one approaches the inner region of the accretion disk. The gas producing the high-velocity wings of lines is exposed to extreme radiation fields that are orders of magnitude greater than what the bulk of the BLR experiences.  There has long been evidence that this produces flatter Balmer decrements. Notably, \citet{Shuder82} discovered a decrease in the H$\upalpha$/H$\upbeta$ ratio in the high-velocity wings of the lines arising from the gas exposed to the most intense continuum.  The Ly$\upalpha$/H$\upbeta$ ratio also shows a velocity dependence.  \citet{Zheng92} found that the Ly$\upalpha$/H$\upbeta$ ratio increased in the high-velocity wings of broad lines. The physical cause of this is discussed in \citet{Gaskell17}.  

However, the bulk of the BLR Balmer line emission does {\em not} come from the high-velocity innermost gas.  Therefore, velocity-integrated ratios very different from Case B are not expected. This argues for changes in intrinsic conditions not having a large effect on object-to-object differences in velocity-integrated hydrogen-line ratios.

A key point of G23 and the present study is that the reddenings implied by the broad hydrogen lines for NGC~5548 and NGC~7469 are similar to those given by the non-hydrogenic reddening indicators (see Table 2 in G23 and Table 1 here).  This supports the intrinsic, velocity-integrated intensity ratios for broad hydrogen lines in AGNs being close to Case B \citep{Gaskell17}, and the observed ratios primarily being a function of reddening. Independent support for this for a larger sample of AGNs comes from the correlation of observed Balmer decrements with observed continuum slopes  \citep{Cackett+07,Heard+Gaskell23}. 


When we are considering temporal variability, the only thing that can drive changes in velocity-integrated line ratios fast enough is changes in the flux of photons heating and ionizing the gas. 
It can be seen in Fig.~4 that the gradual, modest increase in the H$\upalpha$/H$\upbeta$ ratio is not correlated with the faster variations in the UV, even though the individual Balmer line fluxes closely follow the changes in the UV continuum (see Figs.~3 and 4 of \citealt{Collier+98}). Neither is there a correlation with the 2-10 keV X-ray flux.   The lack of correlations with the continua points to the slight gradual change in the observed Balmer decrement being due neither to intrinsic changes in the BLR hydrogen line ratios nor to systematic flux-dependent effects in measuring the hydrogen line intensities.

From the ratio of Ly$\upalpha$ of \citet{Kriss+00} to both H$\upbeta$ and H$\upalpha$ of \citet{Collier+98}, one sees in the two panels of Fig.~5 that both ratios decrease over the monitoring period. As with H$\upalpha$/H$\upbeta$, the change in the ratios is gradual and does not correlate with the $\uplambda$1315 flux. 

\subsection{Reddening variability during longer-term monitoring}

If the variability in H$\upalpha$/H$\upbeta$ we find during the 1996 monitoring is due to changes in reddening, we would expect changes at other times.  Other observers have published long-term spectroscopic and photometric monitoring outside the 1996 monitoring by the {\it International AGN Watch}. Fig.~6 shows the annual averages of the Balmer decrements implied by image-tube observations of H$\upalpha$/H$\upbeta$ ratios as reported by \citet{Doroshenko+94} from 1971 to 1991. Compared to the average of about 5.0 for H$\upalpha$/H$\upbeta$ ratios collected in 1996 by the {\it International AGN Watch}, the ratio provided by the \citet{Doroshenko+94} data is lower until 1985. The H$\upalpha$/H$\upbeta$ ratios then remain relatively stable until 1986, when a sudden increase can be observed until 1989, at which point the ratio returns back to earlier values. It can be noted that the H$\upalpha$/H$\upbeta$ ratio shows a gradual upward trend before the sharper rise in 1986. 

\citet{Osterbrock77} presented observations made with a modern linear detector, and gave an average observed H$\upalpha$/H$\upbeta$ ratio of 3.88 from spectra taken in 1974-1976. This implies $E(B-V)$ = 0.29, in line with the reddenings implied by the \citet{Doroshenko+94} observations, but significantly lower than our estimate from the 1996 campaign and indeed lower than the reddening given by almost all other indicators (see Table 1).

\begin{figure} 
	\begin{center}
		\includegraphics[width = 0.99\linewidth]{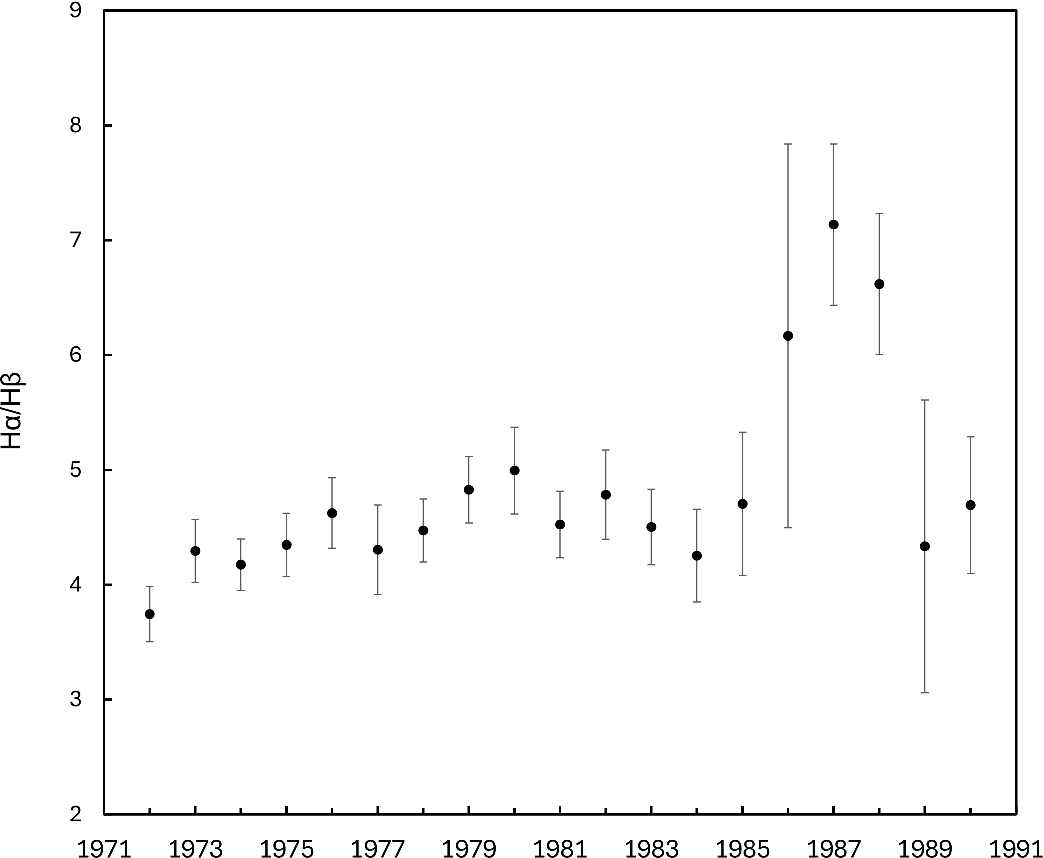}
        \caption{Annual averages of NGC 7469 H$\upalpha$/H$\upbeta$ ratios. Line ratios reported by \citet{Doroshenko+94}. The error bars show errors in the mean.} 
	\end{center}
\end{figure}

\citet{Shapovalova+17} present the results of long-term monitoring from 1996-2015. Unfortunately, their data are inhomogeneous. As they explain, they used different telescopes and apertures (see their Table 2), observed different wavelength regions (see their Table 3), and had different dispersions (see their Section 2.2). In Fig.~7 we show the average reddenings implied by their reported H$\upalpha$/H$\upbeta$ ratios for each observing season with the errors in the means. The season-to-season averages show fluctuations that exceed the errors in the means, but, given the inhomogeneity of the spectra, it is not clear to what extent the fluctuations are real.

Shapovalova et al. also present independent broad-band photometric monitoring from 1996-2015. These observations were made with two photometric telescopes using different apertures (see their Section 2.1 and their Table 4).  To convert the $(B-V)$ colors into $E(B-V)$, it is necessary to know the corrections for host galaxy starlight and to assume an unreddened flux ratio for the $B$ and $V$ bands for the AGN. We have adopted $B$- and $V$-band host galaxy fluxes of 13.75 and 7.6 mJy, and assumed an intrinsic ratio of the $B$ and $V$ fluxes. Because of the uncertainties in the assumptions, there is a systematic uncertainty of $\pm 0.2$ in the $E(B-V)$ values from the photometry. However, we are only concerned with changes from season to season in $E(B-V)$. In Fig.~8, it can be seen that most of the time, season-to-season changes are smaller than the errors in the seasonal means, but sometimes there are larger changes. 

Although, as noted above, there are problems with inhomogeneity for both the spectroscopy and photometry reported by \citet{Shapovalova+17}, the spectroscopy and photometry do have the advantage of being independent (different telescopes, techniques, and nights). If changes in the estimated reddenings are real, we would expect a correlation between the $E(B-V)$ from the photometry and the spectroscopy. In Fig.~9, we show the reddenings implied by the \citet{Shapovalova+17} H$\upalpha$/H$\upbeta$ ratios compared to the reddenings found by the photometric data within $\pm 20$ days. Although the scatter is large, there is a correlation. The correlation coefficient is $r = 0.422$ which has a one-tailed significance of $p = 0.0019$ for 42 degrees of freedom. This supports some of the variation in derived $E(B-V)$ values being real. 

\begin{figure} 
	\begin{center}
		\includegraphics[width = 0.99\linewidth]{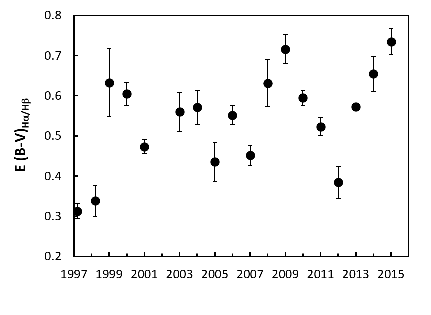}
        \caption{Seasonal average reddenings implied by the H$\upalpha$/H$\upbeta$ ratios reported by \citet{Shapovalova+17} for the period 1996 to 2015. The error bars show the errors in the mean for each season.} 
	\end{center}
\end{figure}

\begin{figure} 
	\begin{center}
		\includegraphics[width = 0.99\linewidth]{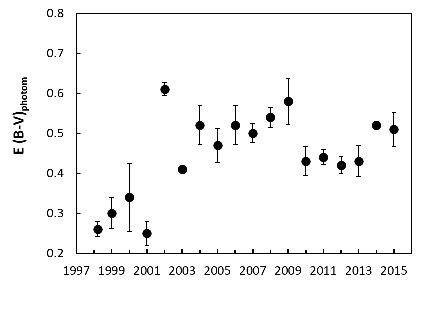}
        \caption{Seasonal average reddenings implied by the $B$ and $V$ broad-band photometry reported by \citet{Shapovalova+17} for the period 1998 to 2015. The error bars again show the errors in the mean for each season.} 
	\end{center}
\end{figure}

\begin{figure} 
	\begin{center}
		\includegraphics[width = 0.99\linewidth]{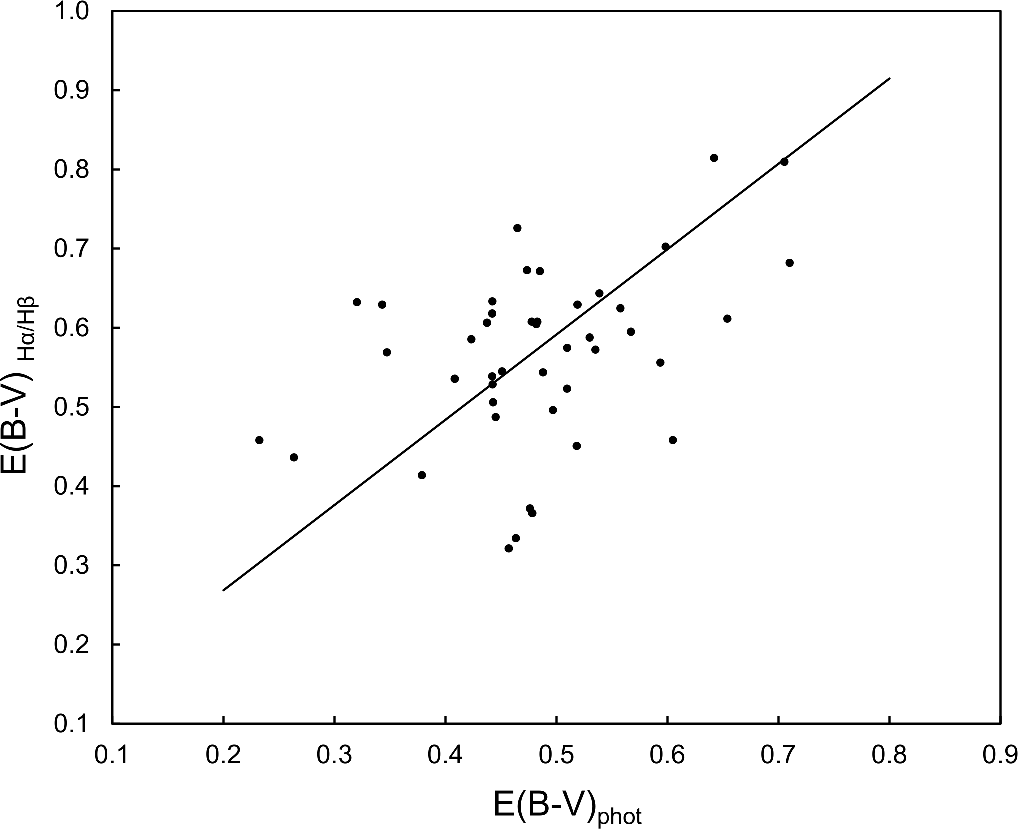}
        \caption{Reddenings estimated from the H$\upalpha$ and H$\upbeta$ fluxes of \citet{Shapovalova+17} versus the reddenings estimated from their broad-band $V$ and $V$ photometry within $\pm 20$ days. The line is a least-square bisector fit.} 
	\end{center}
\end{figure}


\section{Discussion}

\subsection{Variable extinction}

While the gradual changes in hydrogen line ratios during the monitoring period are most easily explained as a gradual increase in extinction, we cannot get the timescale of the change in extinction from the 1996 campaign alone.  However, if the 1984-1991 image-tube observations of \citet{Doroshenko+94} are correct (see Fig.~6), the timescale for a dust cloud to cross the Balmer-line-emitting BLR in NGC~7469 is of the order of 4 years.  This is only slightly longer than the duration of the well-resolved X-ray occultation event in NGC~3227 (see Fig.~3 of \citealt{Lamer+03}). This similarity makes it plausible that the 1985-1990 event in NGC~7469 was due to a dust cloud crossing the line of sight.

From reverberation mapping, the diameter of the BLR producing the Balmer lines is $\sim 12$ light days.  To cross this distance in $\sim 4$ years requires a transverse velocity of $\sim 2500$ km~s$^{-1}$.  This is less than the transverse velocity at the outer edge of the BLR.  \citet{Gaskell+Harrington18} show how partial coverage of the BLR by dust clouds can readily produce the asymmetric Balmer line profiles commonly seen in AGNs and their variability.

\subsection {Accretion disk sizes}

There has been a long-standing problem of observational estimates of accretion disk sizes being around a factor of two larger than predicted by theory. \citet{Gaskell17} proposed that this discrepancy is due to the luminosity of the accretion disk being substantially underestimated because of neglecting internal extinction.  For NGC~7469, our mean reddening of $E(B-V) = 0.44$ corresponds to a V-band extinction of 1.4 magnitudes or a factor of 3.6, which would make the accretion disk 1.9 times bigger.


\section{Conclusions}

We have made estimates of the total Galactic plus internal reddening of NGC 7469 in 1996. This is now the second AGN for which multiple reddening indicators have been used to estimate the reddening.  As found by G23 for NGC~5548, the reddenings given by the different indicators for NGC~7469 are consistent within the errors.  Taken together, they give a mean reddening of $E(B-V) = 0.44 \pm 0.03$ for NGC~5548, which is substantially greater than the reddening of $E(B-V) = 0.017$ due to foreground dust in the Milky Way. 

Our reddening of NGC~7469 and the previously-reported reddening of NGC~5548 imply that there are serious errors in the many studies of AGNs which assume that the only reddening of an AGN is due to foreground dust in the Milky Way and that the internal reddening of AGNs is negligible. Our finding that two well-studied AGNs hitherto widely considered to have little or no internal reddening actually have substantial reddening supports other studies implying that the vast majority of AGNs have substantial reddening.

The large reddening of NGC~7469 provides strong support for the explanation of \citet{Gaskell17} that the discrepancy between predictions and measurements of AGN accretion disk sizes is due to underestimating AGN luminosities by neglecting internal reddening.

As for NGC~5548, the reddenings deduced from the hydrogen lines are consistent with the non-hydrogenic reddening indicators.  This supports the use of the easily-measured H$\upalpha$/H$\upbeta$ ratio as a reddening indicator.

We find that during the 1996 multi-wavelength monitoring campaign the Ly$\upalpha$/H$\upbeta$/Ha ratios gradually changed in a manner that was consistent with a slight increase in reddening. These  changes showed no correlations with the optical/UV variability or the X-ray variability.

Analysis of long-term but less homogeneous spectroscopy before and after 1996 provides some support for there being variations in the extinction in other years. 

We argue that the most plausible explanation of variability in hydrogen line ratios is variable extinction due to dust close to the BLR moving across our line of sight.

 
\begin{acknowledgments}
ARK, EIK and DGS carried out their work under the auspices of the Science Internship Program (SIP) of the University of California at Santa Cruz.  We wish to express our appreciation to Raja GuhaThakurta for his excellent leadership of this program. 
\end{acknowledgments}


\bibliographystyle{aasjournalv7.1}

\end{document}